# Optimized quantum sensing with a single electron spin using real-time adaptive measurements

C. Bonato[1,†], M.S. Blok[1,†], H. T. Dinani[2,3], D. W. Berry[2], M. L. Markham[4], D. J. Twitchen[4], R. Hanson[1,*]

[1]QuTech and Kavli Institute of Nanoscience, Delft University of Technology,
PO Box 5046, 2600 GA Delft, The Netherlands

[2] Department of Physics and Astronomy, Macquarie University,
Sydney, New South Wales 2109, Australia

[3]Center for Engineered Quantum Systems, Macquarie University,
Sydney, New South Wales 2109, Australia

[4]Element Six Ltd, Kings Ride Park, Ascot, Berkshire SL5 8BP, UK,

[†]These authors contributed equally to this work

* corresponding author: r.hanson@tudelft.nl

**Quantum sensors based on single solid-state spins promise a unique combination of sensitivity and spatial resolution[1-20]. The key challenge in sensing is to achieve minimum estimation uncertainty within a given time and with a high dynamic range. Adaptive strategies have been proposed to achieve optimal performance but their implementation in solid-state systems has been hindered by the demanding experimental requirements. Here we realize adaptive d.c. sensing by combining single-shot readout of an electron spin in diamond with fast feedback. By adapting the spin readout basis in real time based on previous outcomes we demonstrate a sensitivity in Ramsey interferometry surpassing the standard measurement limit. Furthermore, we find by simulations and experiments that adaptive protocols offer a distinctive advantage over the best-known non-adaptive protocols when overhead and limited estimation time are taken into account. Using an optimized adaptive protocol we achieve a magnetic field sensitivity of 6.1 ± 1.7 nT Hz$^{-1/2}$ over a wide range of 1.78 mT. These results open up a new class of experiments for solid-state sensors in which real-time knowledge of the measurement history is exploited to obtain optimal performance.**

Quantum sensors have the potential to achieve unprecedented sensitivity by exploiting control over individual quantum systems[1,2]. As a prominent example, sensors based on single electron spins associated with Nitrogen-Vacancy (NV) centers in diamond capitalize on the spin's quantum coherence and the high spatial resolution resulting from the atomic-like electronic wave function[3,4]. Pioneering experiments have already demonstrated single-spin sensing of magnetic fields[5-7], electric fields[8], temperature[9,10] and strain[11]. NV sensors may therefore have a revolutionary impact on biology[12-15], nanotechnology[16-18] and material science[19,20].

A spin-based magnetometer can sense a d.c. magnetic field $B$ through the Zeeman shift $E_z = \hbar \gamma B = \hbar 2\pi f_B$ ($\gamma$ is the gyromagnetic ratio and $f_B$ the Larmor frequency) between two spin levels $|0\rangle$ and $|1\rangle$. In a Ramsey interferometry experiment, a superposition state $(1/\sqrt{2})(|0\rangle + |1\rangle)$, prepared by a π/2 pulse, will evolve to $(1/\sqrt{2})(|0\rangle + e^{i\varphi}|1\rangle)$ over a sensing time $t$. The phase $\varphi = 2\pi f_B t$ can be measured by reading out the spin in a suitable basis, by adjusting the phase $\vartheta$ of a second π/2 pulse.

For a Ramsey experiment that is repeated with constant sensing time $t$ the uncertainty $\sigma_{f_B}$ decreases with the total sensing time $T$ as $1/(2\pi\sqrt{tT})$ (standard measurement sensitivity, SMS). However, the field range also decreases with $t$ because the signal is periodic, creating ambiguity whenever $|2\pi f_B t| > \pi$. This results in a dynamic range bounded as $f_{B,max}/\sigma_{f_B} \leq \pi\sqrt{T/t}$. Recently, it was discovered that the use of multiple sensing times within an estimation sequence can yield a scaling of $\sigma_{f_B}$ proportional to $1/T$, resulting in a vastly improved dynamic range: $f_{B,max}/\sigma_{f_B} \leq$

$\pi T/\tau_{min}$, where $\tau_{min}$ is the shortest sensing time used. A major open question is whether adaptive protocols, in which the readout basis is optimized in real time based on previous outcomes, can outperform non-adaptive protocols. While scaling beating the standard measurement limit has been reported with non-adaptive protocols[22,23], feedback techniques have only recently been demonstrated for solid-state quantum systems[24-26] and adaptive sensing protocols have so far remained out of reach.

Here we implement adaptive d.c. sensing with a single-electron spin magnetometer in diamond by exploiting high-fidelity single-shot readout and fast feedback electronics (Fig. 1a). We demonstrate a sensitivity beyond the standard measurement limit over a large field range. Furthermore, we investigate through experiments and simulations the performance of different adaptive protocols and compare these to the best known non-adaptive protocol. Although the non-adaptive protocol improves on the standard measurement limit for sequences with many detections we find that the adaptive protocols perform better when overhead time for initialization and readout is taken into account. In particular, the adaptive protocols require shorter sequences to reach the same sensitivity, thus allowing for sensing of signals that fluctuate on shorter timescales.

Our magnetometer employs two spin levels of a single NV center electron in isotopically purified diamond (0.01% $^{13}$C). We exploit resonant spin-selective optical excitation, at a temperature of 8 K, for high-fidelity initialization and single-shot readout[27] (Fig. 1b). Microwave pulses, applied via an on-chip stripline, coherently control the electron spin state. From Ramsey experiments, we measure a spin dephasing time of $T_2^* = (96 \pm 2)$ μs (Fig. 1c). In order to characterize the performance of different sensing protocols in a controlled setting, the effect of the external field is implemented as an artificial frequency detuning, by adding $\varphi = 2\pi f_B t$ to the phase $\vartheta$ of the final π/2 pulse.

To achieve high sensitivity in a wide field range we use an estimation sequence consisting of $N$ different sensing times[21-23,28], varying as $\tau_n = 2^{N-n}\tau_{min}$ ($n = 1..N$). The value of $\tau_{min}$ sets the range; we take $\tau_{min}$= 20 ns, corresponding to a range $|f_B| < 25$ MHz, equivalent to $|B|$<0.89 mT for $\gamma = 2\pi \cdot 28$ MHz mT$^{-1}$.

The key idea of adaptive magnetometry is that for each Ramsey experiment the measurement basis is chosen based on the previous measurement outcomes such that the uncertainty in the frequency estimation is minimized (Fig. 1a). After every Ramsey experiment, the outcome is used to update a frequency probability distribution $P(f_B)$ according to Bayes' rule, taking measured values for detection fidelity and coherence time into account (Methods). The current estimate of $P(f_B)$ is then used to calculate the phase $\vartheta$ of the final π/2 pulse which allows for best discrimination between different possible magnetic field values in the next Ramsey experiment[28]. In our experiment, this process is realized by a microprocessor, which receives the measurement outcome, performs the Bayesian estimate, calculates the phase $\vartheta$ and subsequently sends a digital signal to a field-programmable gate array (FPGA) to adjust the phase of the final π/2 pulse accordingly (Fig 1a).

To reduce the undesired effects of quantum projection noise and imperfect readout fidelity we perform $M_n$ Ramsey experiments[21] for each sensing time $\tau_n$, with $M_n = G + F(n - 1)$. Here $G$ and $F$ can be chosen to optimize the performance of the protocol. For the short sensing times (large $n$), corresponding to the measurements that make the largest distinction in frequency (and where an error is therefore most detrimental), we perform the most Ramsey experiments. We will compare several protocols that differ in the strategy of adaptive phase choice. As a first example, we consider a protocol where the phase $\vartheta$ is adjusted each time the sensing time is changed; we name this "limited-adaptive" protocol.

An example of the working principles of the limited-adaptive protocol is illustrated in Fig. 2, for an estimation sequence comprising $N = 3$ sensing times and one measurement per sensing time ($G = 1, F = 0$). We start with no information over $f_B$, corresponding to a uniform probability density $P(f_B)$ (solid black line). For the first Ramsey experiment, the sensing time is set to $4\tau_{min}$. $P(f_B)$ is updated depending on the measurement outcome. For example, the outcome 1 indicates maximum probability for the values $f_B = \pm 6.25, \pm 18.75$ MHz, and minimum probability for $f_B = 0, \pm 12.5, \pm 25$ MHz. This indeterminacy in the estimation originates from the fact that, for this sensing time, the acquired phase spans the range [-4π, 4π], wrapping multiple times around the [-π, π] interval. The sensing time is then decreased to $2\tau_{min}$. Given the current $P(f_B)$ for outcome 1 (black curve), the filter functions that would be applied to $P(f_B)$ after the Bayesian update for detection outcomes 0 and 1 are represented, respectively, by the light red and blue areas. For $\vartheta = -\pi/2$, maximum distinguishability is ensured: outcome 0 would select the peaks around $f_B = -6.25, +18.75$ MHz, while outcome 1 would select the peaks around $f_B = -18.75, +6.25$ MHz. The same process is then repeated, decreasing the sensing time to $\tau_{min}$. The remaining uncertainty, corresponding to the width of the resulting peak in $P(f_B)$, is set by the longest sensing time $4\tau_{min}$.

Figure 3b shows the probability density yielded by experimental runs of the limited-adaptive protocol with different numbers of sensing times $N$ = 1,3,5,7,9. Here, $f_B = 2$ MHz, and each estimation sequence is repeated 101 times, with $G = 5, F = 7$. For increasing $N$, the width of the distribution becomes more narrowly peaked around the expected value of 2 MHz, while the wings of distribution are strongly suppressed.

To verify that the protocol works over a large dynamic range, we measure the uncertainty as a function of detuning $f_B$. To account for the periodic nature of phase we use the Holevo variance $V_H = (|\langle e^{i2\pi f_B^{est}\tau_{min}}\rangle|)^{-2} - 1$ as a measure of the uncertainty. We estimate $f_B^{est}$ by taking the mean of the probability density $P(f_B)$ resulting from a single run of the protocol. A fixed initial phase ($\vartheta = 0$ in our experiments) results in a specific dependence of the variance on the magnetic field. For example, for $N$ = 2, only four measurement outcomes are possible {00, 01, 10, 11}, corresponding to $f_B = 0$, -25, -12.5, +12.5 MHz, respectively. These specific detunings can be measured with the highest accuracy since they correspond to measurements of an eigenstate of our quantum sensor at the end of the Ramsey experiment, while for other frequencies larger statistical fluctuations will be found due to spin projection noise. Figure 3c shows $V_H$ as a function of detuning for the parameters $G$=5, $F$=7. Both the experimental data (dots) and the numerical simulation (solid lines) confirm the expected periodic behavior.

We now use our adaptive sensing toolbox to compare different sensing protocols by investigating the scaling of $\eta^2 = V_H T$, averaged over different detunings, as a function of the total sensing time $T$. First, we will ignore the overhead time due to spin initialization and readout.

We compare the limited-adaptive protocol to the best known non-adaptive protocol and to an optimized adaptive protocol. In the non-adaptive protocol[21-23], the readout phase for the $m$-th Ramsey experiment is always set to $\vartheta_{n,m} = \frac{m\pi}{M_n}$ ($m = 1..M_n$). In the optimized adaptive protocol[29,30], the phase $\vartheta$ is updated before each Ramsey and, additionally, a phase $\vartheta_{n,m}^{incr}$ that depends only on the current values of $n, m$ and the last measurement outcome $\mu_{n,m}$, is added. This additional phase $\vartheta_{n,m}^{incr}$ is determined by a numerical minimization of the Holevo variance, via swarm-optimization techniques, taking experimental parameters into account. A detailed description of all protocols is reported in Supplementary Tables 1-3.

We plot experimental data for the sensitivity scaling for the three protocols in Fig. 4a alongside simulations using known experimental parameters. In these graph, the SMS limit corresponds to a constant $V_H T$; any scaling behavior with a negative slope thus improves beyond the SMS.

We observe that, for the setting ($G$=5, $F$=2), the non-adaptive protocol reaches only the SMS limit, while both adaptive protocols yield $V_H T$ scaling close to 1/$T$. When the number of measurements per interaction time is increased to ($G$=5, $F$=7) the non-adaptive protocol also shows sub-SMS scaling (Fig. 4a, blue line). We find this behavior to be quite general: both adaptive and non-adaptive protocols can reach 1/$T$ scaling, but the adaptive protocols require fewer measurements (Supplementary Figures S1-S4). By comparing the best non-adaptive and the best adaptive protocol, we find that they reach the same sensitivity of (6.1 ± 1.7) nT Hz$^{-1/2}$ when the longest sensing time reaches $T_2^*$. The non-adaptive protocol however, requires significantly more measurements (611) than the adaptive protocol (221).

The advantage of adaptive measurements becomes clear when the initialization and readout times (overhead) are taken into account (Fig 4b). Since the time required to compute the controlled phase is similar to the initialization time, the two operations can be performed simultaneously, with no additional overhead required by the adaptive protocol (Methods). While the two best protocols still achieve similar minimum sensitivities, the optimized adaptive protocol requires significantly less measurement time. At any fixed measurement time, the adaptive protocol estimates the magnetic field more accurately, allowing a higher repetition rate for the estimation sequences. This is advantageous in the realistic situation that the magnetic field to be estimated is not static: in this case, the estimation time is required to be shorter than the timescale of the fluctuations. Our data shows that at an estimation repetition rate of 20 Hz, the non-adaptive protocol can estimate a magnetic field with an sensitivity $\eta$ = (749 ± 35) nT Hz$^{-1/2}$, while the optimized-adaptive protocol yields $\eta$ = (47 ± 2) nT Hz$^{-1/2}$.

While the record sensitivity reported here is enabled by single-shot spin readout at low temperature, adaptive techniques can prove valuable also in experiments at room temperature[23] where spin-dependent luminescence intensity under off-resonant excitation is typically used to measure the electronic spin. By averaging the signal over multiple repetitions an arbitrarily high readout fidelity can be achieved ($F$=0.99 for 50,000 repetitions[23]). Interestingly, we find that the benefits given by adaptive techniques persist also in case of lower readout fidelities and that the combination of adaptive techniques and optimization of the number of readout repetitions yields a significant improvement (Supplementary Figure S6).

In conclusion, by combining high-fidelity single-shot readout at low temperature with a single electron spin sensor and fast electronics, we achieve an unprecedented d.c. sensitivity of (6.1±1.7) nT Hz$^{-1/2}$ with a repetition rate of 20 Hz. Another relevant figure of merit for sensors is the ratio between the range and the sensitivity; the best value found in this work ($B_{max}/\eta \sim 1.5 \cdot 10^5$ Hz$^{1/2}$) improves on previous experiments by two orders of magnitude[22,23]. Furthermore, we found that the best known adaptive protocol outperforms the best known non-adaptive protocol when overhead is taken into account. These insights can be extended to other quantum sensors and to the detection of different physical quantities such as temperature and electric fields. A remaining open question is whether this adaptive protocol is optimal; perhaps further improvements are possible by taking into account the full measurement history. In a more general picture, the adaptive sensing toolbox demonstrated in this work will enable exploration of the ultimate limits of quantum metrology and may lead to practical sensing devices combining high spatial resolution, sensitivity, dynamic range and repetition rate.

**METHODS**

1. **Sample and experimental setup**

    We use an isotopically-purified diamond sample, grown by Element Six Ltd., with 0.01% $^{13}$C content. Experiments are performed in a flow cryostat, at the temperature of 8 K. A magnetic field of 12 Gauss is applied to split the energies of the $m_s = \pm 1$ spin states, in order to provide selective spin control by resonant microwave driving. A solid immersion lens is fabricated on top of the NV center by focused ion beam, and covered with an anti-reflective layer, to increase photon collection efficiency.

    The experiment is controlled by an Adwin Gold microprocessor, with 1 MHz clock cycle. The microprocessor updates the frequency estimate based on the measurement outcomes and calculates the controlled phase. The phase is then converted to an 8-bit number, sent to the FPGA. The FPGA outputs an IQ modulated, 30 MHz sinusoidal pulse, with the specified controlled phase, which drives a vector microwave source.

2. **Adaptive algorithm**

    For the $\ell$-th Ramsey experiment, with outcome $\mu_\ell$ (0 or 1), the estimate of the magnetic field is updated according to Bayes rule: $P(f_B|\mu_1 \ldots \mu_\ell) \sim P(f_B|\mu_1 \ldots \mu_{\ell-1})P(\mu_\ell|f_B)$, with a normalizing proportionality factor. $P(\mu_\ell|f_B)$ is the conditional probability of outcome $\mu_\ell$ (0 or 1) given a frequency $f_B$:

$$P(\mu = 0|f_B) = \frac{(1 + F_0 - F_1)}{2} + \frac{(F_0 + F_1 - 1)}{2} e^{-\left(\frac{t}{T_2^*}\right)^2} \cos[2\pi f_B t + \vartheta]$$

$$P(\mu=1|f_B) = 1 - P(\mu=0|f_B)$$

where $t = 2^{N-n}\tau_{min}$. Due to its periodicity, it is convenient to express $P(\mu|f_B)$ in a Fourier series, resulting in the following update rule:

$$p_k^{(\ell)} = \frac{1+(-1)^{\mu_\ell}(F_0-F_1)}{2}p_k^{(\ell-1)}$$
$$+ e^{-\left(\frac{\tau}{T_2^*}\right)^2}\frac{(F_0+F_1)-1}{4}\left[e^{i(\mu_\ell\pi+\vartheta_\ell)}p_{k-2^{N-n}}^{(\ell-1)} + e^{-i(\mu_\ell\pi+\vartheta_\ell)}p_{k+2^{N-n}}^{(\ell-1)}\right]$$

The Bayesian update is performed using the experimental values $F_0$ = 0.88, $F_1$ = 0.98 and $T_2^*$ = 96 µs.

The Holevo variance after each detection, expressed as $V_H = (2\pi|p_{2^{N-n+1}}^{(\ell-1)}|)^{-2} - 1$, can be minimized by choosing, at each step, the following controlled phase for the second π/2 pulse[28]:

$$\vartheta^{ctrl} = \frac{1}{2}\arg\left\{p_{2^{N-n+1}}^{(\ell-1)}\right\}$$

In the limited-adaptive protocol, this phase is recalculated every time the sensing time is changed. For the optimized-adaptive protocol, the controlled phase is recalculated before every Ramsey experiment and the phase of the second π/2 pulse is set to $\vartheta = \vartheta_{\ell,m}^{ctrl} + \vartheta_{n,m}^{incr}$, where $\vartheta_{n,m}^{incr}$ is a phase increment that depends on the last measurement outcome[30].

To avoid exceeding the memory bounds of the microprocessor, and to optimize speed, we need to minimize the number of coefficients to be tracked and stored. This can be done by determining which coefficients are non-zero and contribute to $p_{2^{N-n+1}}^{(\ell-1)}$ and neglecting the rest. Moreover, since the probability distribution is real, $\left(p_k^{(\ell)}\right)^* = p_{-k}^{(\ell)}$; therefore we only store and process coefficients $p_k^{(\ell)}$ with $k>0$.

For each Ramsey run, in the case ($G=5, F=2$), the time taken by the microprocessor to perform the Bayesian update ranges between 80µs and 190µs. This time is comparable to the spin initialization duration, so both operations can be performed simultaneously, with no additional overhead (Supplementary Table 4).

## Acknowledgements

We thank Marijn Tiggelman and Raymond Schouten for the development of the FPGA. We acknowledge support from the Dutch Organization for Fundamental Research on Matter (FOM), the Netherlands Organization for Scientific Research (NWO), the DARPA QuASAR programme, the EU SOLID, and DIAMANT programmes and the European Research Council through a Starting Grant. D.W.B. is funded by an Australian Research Council Future Fellowship (FT100100761).

Fig. 1

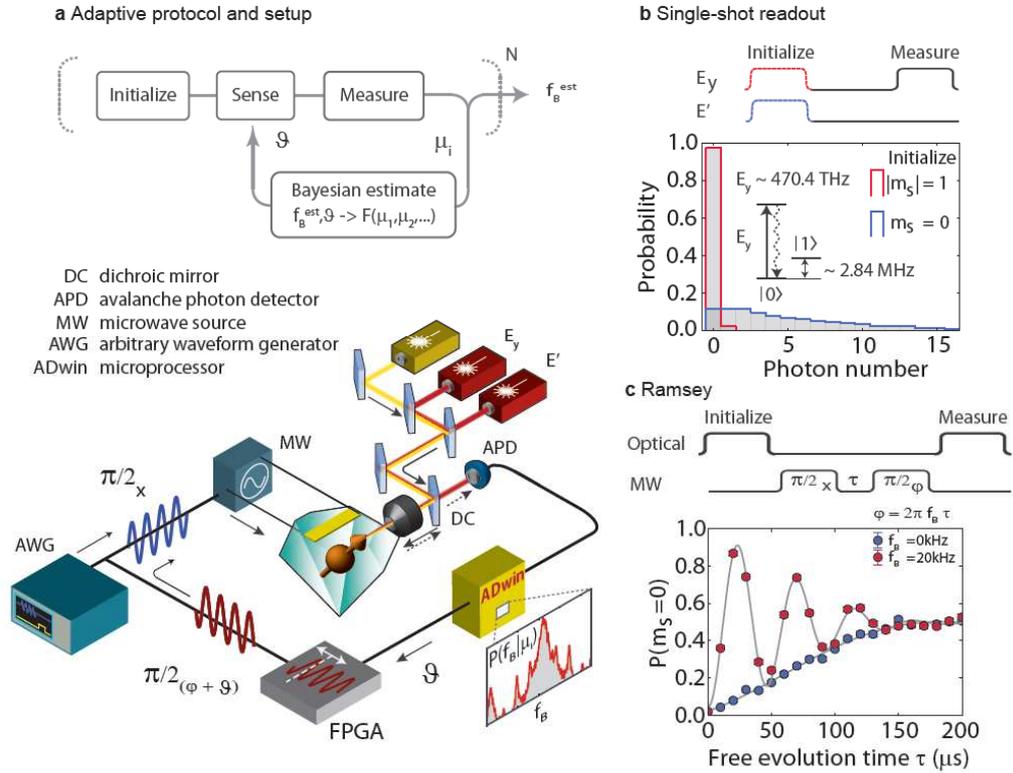

**Figure 1. Experiment concept and apparatus.** (a) The adaptive frequency estimation protocol consists of a sequence of initialization, sensing, measurement operations. After each measurement run, the outcome $\mu$ is used to update the estimate of the frequency $f_B$, which is then used to optimize the sensing parameters for the following run. Experimentally, the frequency estimation and adaptive calculation of the phase are performed in real-time by a microprocessor. (b) The experiment is performed using the states $|0\rangle = |m_s = 0\rangle, |1\rangle = |m_s = -1\rangle$ of the electronic spin of a NV centre in diamond. The electronic spin is readout by resonant optical excitation and photon counting[27]: detection of luminescence photons corresponds to detection of the $|0\rangle$ state. We plot the probability of detecting a photon after initializing either in $|0\rangle$ or $|1\rangle$. The readout fidelities for the states $|0\rangle$ (outcome 0) and $|1\rangle$ (outcome 1) are $F_0 = 0.88 \pm 0.02$, $F_1 = 0.98 \pm 0.02$, respectively. (c) Each measurement run consists of a Ramsey experiment, in which the phase accumulated over time by a spin superposition during free evolution is measured. The measurement basis rotation is controlled by the phase $\vartheta$ of the final π/2 pulse. From the measured phase, we can extract the frequency $f_B$, corresponding to an energy shift between the levels $|0\rangle$ and $|1\rangle$ given by an external field (magnetic field, temperature, strain...). Here, to test the performance of different protocols, we set $f_B$ as an artificial detuning, set by the microprocessor by adding $\varphi = 2\pi f_B t$ to the phase $\vartheta$ (Supplementary Figure S7).

## Fig. 2 High dynamic range adaptive magnetometry

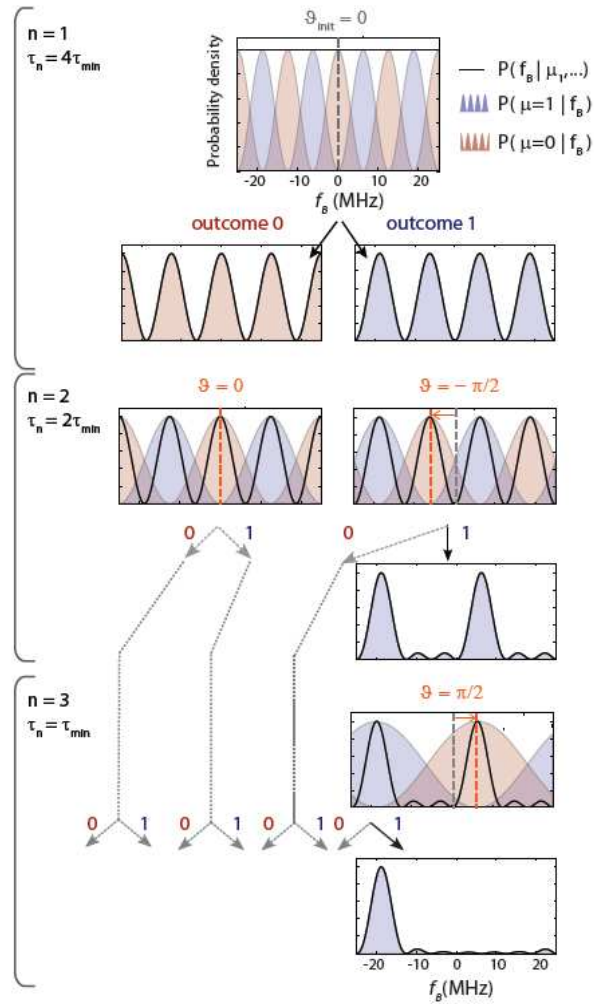

**Figure 2. High dynamic-range adaptive magnetometry** Limited-adaptive protocol, in the case of one Ramsey experiment per sensing time (*G*=1, *F*=0). In each step, the current frequency probability distribution $P(f_B)$ is plotted (solid black line), together with conditional probabilities $P(\mu|f_B)$ for the measurement outcomes $\mu = 0$ (red shaded area) and $\mu = 1$ (blue shaded area). After each measurement, $P(f_B)$ is updated according to Bayes' rule. The detection phase $\vartheta$ of the Ramsey experiment is set to the angle which attains the best distinguishability between peaks in the current frequency probability distribution $P(f_B)$. Ultimately, the protocol converges to a single peak in the probability distribution, which delivers the frequency estimate.

Fig. 3

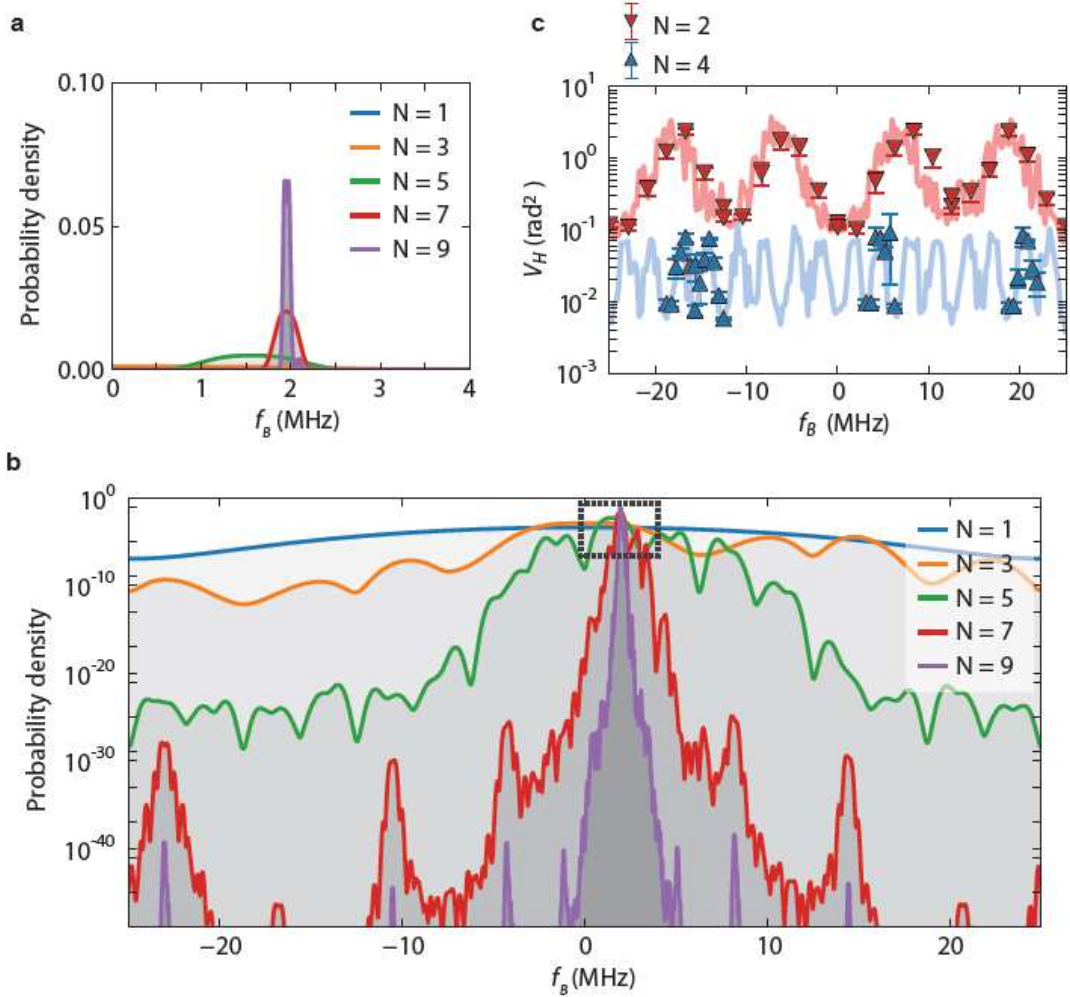

**Figure 3. Frequency dependence of uncertainty.** (a)-(b) Frequency estimate example, for ($G$=5, $F$=7). We set a fixed artificial detuning $f_B = 2$ MHz and run different instances of the limited-adaptive frequency estimation protocol, with increasing $N$. The resulting probability density $P(f_B)$ is averaged over 101 repetitions. (c) Holevo variance as a function of the frequency $f_B$ for $N$=2, 4 (limited-adaptive protocol, $G$=5, $F$=7). We vary $f_B$ by adjusting the phase of the final $\pi/2$ pulse. Solid lines correspond to numerical simulations, performed with 101 repetitions per frequency point and experimental parameters for fidelity and dephasing. Experimental points (triangular shape), were acquired with 101 repetitions each. Error bars (one standard deviation) are calculated by bootstrap analysis.

# Fig. 4 Scaling

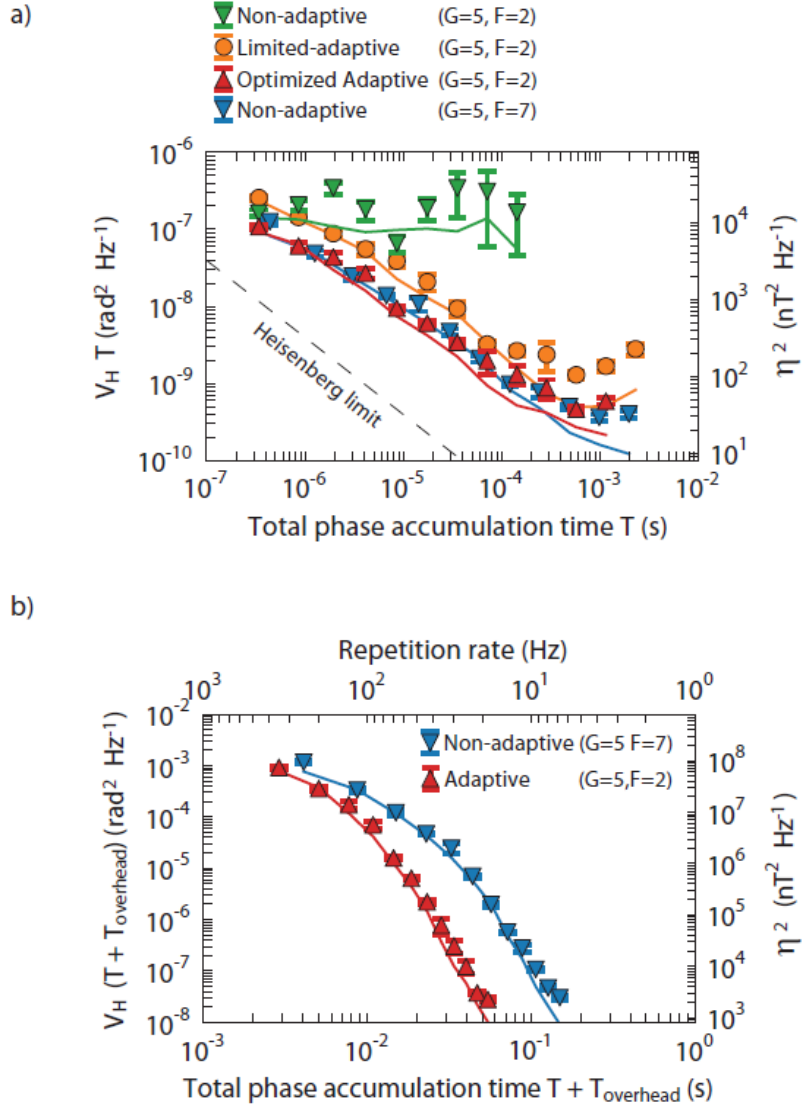

**Figure 4. Scaling of sensitivity as a function of total time.** (a) The three protocols are compared by plotting $\eta^2 = V_H T$ as a function of the total sensing time $T$ (not including spin initialization and readout). For ($G=5$, $F=2$) the non-adaptive protocol (green triangles) is bound to the SMS limit, while for both the limited-adaptive (orange circles) and the optimized adaptive (red triangles) protocols $\eta^2$ scales close to $1/T$. The sensitivity of the limited-adaptive protocol is, however, worse than the optimized-adaptive one. When increasing the number of Ramsey experiments per sensing time to ($G=5$, $F=7$), the non-adaptive protocol (blue triangles) reaches Heisenberg-like scaling, with a sensitivity comparable to the optimized adaptive protocol for ($G=5$, $F=2$). (b) By including spin initialization and readout durations, the superiority of the optimized adaptive protocol (red triangles), which requires less Ramsey runs per sensing time (smaller $F$, $G$) to reach $1/T$ scaling, is evidenced. The optimized adaptive protocol can estimate magnetic fields with a repetition rate of 20Hz, with a sensitivity more than one order of magnitude better than the non-adaptive protocol.

All data are taken with 700 repetitions per data-point. In both plots, error bars corresponding to one standard deviation of the results are obtained using the bootstrap method.

# Supplementary Information

# Optimized quantum sensing with a single electron spin using real-time adaptive measurements

C. Bonato, M.S. Blok, H. T. Dinani, D. W. Berry, M. L. Markham, D. J. Twitchen, R. Hanson

## I. Comparison of sensing protocols: numerical simulations

In the following, the performances of different single-qubit frequency estimation protocols will be compared through numerical simulations. We will describe and analyse three main sensing algorithms, defined using a pseudo-code in Supplementary Tables 1, 2, 3: the limited-adaptive, non-adaptive and optimized-adaptive protocols.

In order to achieve high dynamic range, each estimation sequence consists of N different sensing times, multiples of the shortest sensing time $\tau_{min} = 20$ ns: $\tau_n = 2^{N-n}\tau_{min}$ $(n = 1..N)$.

After each Ramsey, the electron spin is measured: the result of each detection is indicated in the pseudo-code by the **Ramsey (ϑ, τ)** function. The input parameters of this function are the sensing time τ and the phase ϑ of the second π/2 pulse. In the simulation code, this function generates a random value $\mu$ ($\mu = 0, 1$), using the python library *numpy.random*, chosen according to the probability distribution [$p_0$, $p_1=1-p_0$], where:

$$p_0 = P(0|f_B) = \frac{(1+F_0-F_1)}{2} + \frac{(F_0+F_1-1)}{2} e^{-\left(\frac{\tau}{T_2^*}\right)^2} \cos[2\pi f_B \tau + \vartheta] \quad \text{(Eq. S-E1)}$$

$F_0, F_1$ are, respectively the readout fidelities for $m_s=0$ and $m_s=1$. In the following simulations we use the values: $F_0 = 0, 0.75, 0.88$ or $1.00$ ($m_S = 0$), $F_1 = 0.993$ ($m_S = 1$), $T_2^* = 5$μs or 96 μs.

For each Ramsey experiment (indexed here by the label $\ell$), the detection result $\mu_\ell$ is used to update the estimation of the magnetic field using Bayes rule: $P(f_B|\mu_1 \ldots \mu_\ell) \sim P(f_B|\mu_1 \ldots \mu_{\ell-1})P(\mu_\ell|f_B)$. This is indicated in the pseudo-code by the function **Bayesian_update (res, ϑ, τ)**.

Due to its periodicity it is convenient to express $P(f_B)$ in a Fourier series, resulting in the following update rule:

$$p_k^{(\ell)} = \frac{1+(-1)^{\mu_\ell}(F_0-F_1)}{2} p_k^{(\ell-1)} + e^{-\left(\frac{\tau}{T_2^*}\right)^2} \frac{(F_0+F_1)-1}{4} \left[e^{i(\mu_\ell\pi+\vartheta_\ell)} p_{k-2^{N-n}}^{(\ell-1)} + e^{-i(\mu_\ell\pi+\vartheta_\ell)} p_{k+2^{N-n}}^{(\ell-1)}\right] \quad \text{(Eq. S-E2)}$$

Given the periodic nature of phase, the uncertainty is better estimated using the Holevo variance $V_H = (|\langle e^{i2\pi f_B \tau_{min}}\rangle|)^{-2} - 1 = (2\pi |p_{2^{N-n+1}}^{(\ell-1)}|)^{-2} - 1$. The Holevo variance can be minimized by choosing the controlled phase [PS-1]:

$$\vartheta^{ctrl} = \frac{1}{2}\arg\left\{p_{2^{N-\ell+1}}^{(\ell-1)}\right\} \quad \text{(Eq. S-E3)}$$

One Ramsey experiment per sensing time does not allow to reach the Heisenberg-like scaling since the resulting probability distribution, despite being strongly peaked around the expected value, has very large wings with non-zero probability of outlier outcomes. Outliers, although occurring infrequently, can significantly alter the estimate statistics. While this is true for perfect readout ($F_0 = F_1 = 1$) the algorithm performance is reduced even further by imperfect readout [PS-2]. A solution to these problems is to perform $M_n$ Ramsey measurements for each interaction time, with $M_n = G + F(n-1)$ [PS-2].

**Supplementary Table 1:
Limited-adaptive protocol**

*for n = 1 to N:*
   $t_n = 2^{N-n}$
   **choose $\vartheta_n^{ctrl}$**
   $M_n = G + F(n-1)$

   *for m = 1 to $M_n$:*
      $\mu$ = Ramsey ($\vartheta = \vartheta_n^{ctrl}$, $\tau = t_n\tau_{min}$)
      Bayesian_update (res =$\mu$, $\vartheta = \vartheta_n^{ctrl}$, $\tau = t_n\tau_{min}$)

*Supplementary Table 2:
Non-adaptive protocol*

*for n = 1 to N:*
   $t_n = 2^{N-n}$
   $M_n = G + F(n-1)$

   *for m = 1 to $M_n$:*
      **$\vartheta_{n,m} = (m-1)\pi/M_n$**
      $\mu$ = Ramsey ($\vartheta = \vartheta_{n,m}$, $\tau = t_n\tau_{min}$)
      Bayesian_update (res = $\mu$, $\vartheta = \vartheta_{n,m}$, $\tau = t_n\tau_{min}$)

For each protocol it is crucial to find the optimal values for $F$ and $G$, given the experimental readout fidelities $F_0$ and $F_1$. The relevant figure of merit is the sensitivity $\eta$, defined as $\eta^2 = V_H T$. Simulations are performed by running the protocol for 315 different values of the frequency $f_B$, over 31 repetitions for each value. The detection phase $\vartheta$ of the Ramsey is initially set to zero.

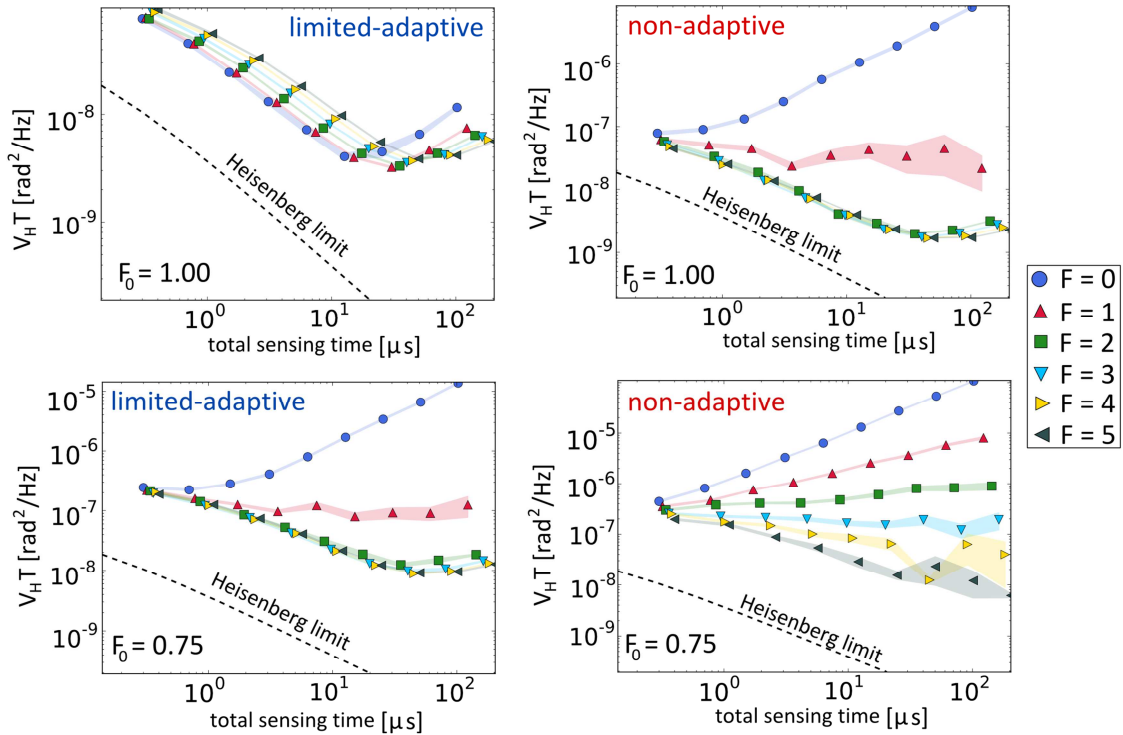

**Supplementary Figure S1.** *Simulations comparing the limited-adaptive and non-adaptive protocols for G = 5, for different values of F, with $T_2^* = 5\mu s$. On the top row, perfect readout fidelity ($F_0 = 1$), on the bottom row, $F_0 = 0.75$. The shaded areas correspond to uncertainties (one standard deviation, calculated by a bootstrap technique). Note that the sensitivity is not further improved after the limit of $T_2^*$ is reached (total sensing time T ~ 10µs).*

**Limited-adaptive vs non-adaptive protocols.** The *limited-adaptive protocol* [PS-1] is described by the pseudo-code in Box-1. The controlled phase $\vartheta^{ctrl}$ is updated every time the sensing time is changed. In this section, the performance of the limited-adaptive protocol will be compared to the *non-adaptive protocol* described by Said et al. [PS-2] and demonstrated experimentally by Waldherr et al.

[PS-4] and Nusran et al. [PS-5]. The pseudo-code for the non-adaptive protocol is reported in Box-2. In this case, the phase of the Ramsey experiment is not updated in real-time based on the estimation of magnetic field given by the previous measurement outcomes, but its value is swept between 0 and π according to predefined values. If, for a given sensing time, $M_n$ Ramsey experiments are performed, the Ramsey phase is increased at each step by $\pi/M_n$.

A comparison of the sensitivity as a function of sensing time $T$ for different values of $F$ (fixing $G = 5$) is shown in Supplementary Figure S1. The data-points correspond to estimation sequences with increasing $N$ ($N = 2..10$). The total sensing time $T$, for each estimation sequence, is calculated as:

$$T = \tau_{min}[G(2^N - 1) + F(2^N - N - 1)] \quad \text{(Eq. S-E4)}$$

In top row of Supplementary Figure S1, the sensitivities for the adaptive and non-adaptive protocols are compared in the case of perfect readout fidelities. In this case, the adaptive protocol follows a Heisenberg-like scaling already for $F = 0$, even though the minimum sensitivity can only be reached for $F = 1$. On the other hand, the non-adaptive protocol requires at least $F = 2$ to reach Heisenberg-like scaling. On the bottom row, we compare the sensitivities for reduced readout fidelity ($F_0 = 0.75$). Here, the adaptive protocol reaches HL-scaling for $F \geq 2$, while the non-adaptive protocol can only get close to it with $F = 5$.

It is important to stress that, in both cases, there is a big improvement when $M_n$ is a function of $n$ ($F > 0$) compared to the case where $M_n$ is independent of $n$ ($F = 0$). In other words, it is beneficial to repeat more often Ramsey experiments with shorter sensing time. The reason is two-fold: on one end they contribute less to the total sensing time, on the other end they are related to larger frequencies which, if estimated wrong, would give a larger error.

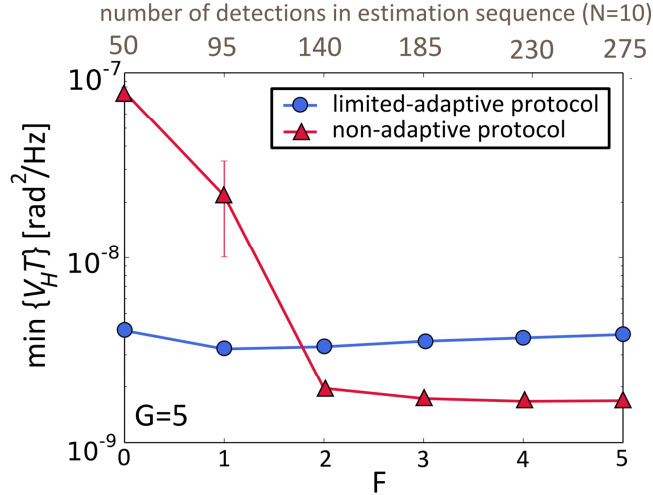

**Supplementary Figure S2.** *Simulation results for the best achieved sensitivity, comparing the limited-adaptive and non-adaptive protocols as a function of $F$. Here, we assume perfect readout fidelity ($F_0 = 1$) and $T_2^* = 5\mu s$. On the top x-axis, the total number of Ramsey experiments in the estimation sequence for $N = 10$ is reported. Error bars, corresponding to one standard deviation, are calculated by bootstrap.*

Comparison between protocols is easier when plotting only the minimum sensitivity vs F. This is shown in Supplementary Figure S2 for perfect readout fidelity $F_0 = 1$. We find that for $F<2$, the limited-adaptive protocol outperforms the non-adaptive protocol. This is expected since in this region only the limited-adaptive protocol exhibits Heisenberg-like scaling. However, once the non-adaptive protocol achieves Heisenberg-like scaling ($F \geq 2$) it reaches a lower sensitivity.

On the scale at the top of Supplementary Figure S2, the number of Ramsey runs corresponding to an estimation sequence with $N = 10$ different sensing times is reported. By increasing $F$, the number of Ramsey experiments increases as:

$$R_N = GN + \frac{FN(N-1)}{2} \qquad \text{(Eq. S-E5)}$$

For perfect readout fidelity, the limited-adaptive protocol reaches HL-scaling for $F = 0$: therefore it only requires $R_N$ = 50 Ramsey runs in the estimation sequence. On the other hand, the non-adaptive requires $F = 2$, i.e. $R_N = 140$ Ramsey runs. Each Ramsey comprises an initialization/measurement duration, labelled as "overhead", not included in the plots (where we only take the sensing time into account). In practice, it is however necessary to minimize the total time of the sequence (including overhead), so that protocols that achieve Heisenberg-like-scaling with smaller $F$ (and therefore less detections $R_N$) are to be preferred as discussed in the main text (Fig. 4).

A striking result is the fact that, once the non-adaptive protocol reaches Heisenberg-like-scaling, it achieves a better sensitivity than the limited-adaptive one. Since non-adaptive protocols are a particular case of the most general class of adaptive protocols, this indicates that the limited-adaptive protocol is not optimal and that protocols with better performance must exist.

**Optimized adaptive protocol.** In order to improve the performance of the limited-adaptive protocol, we consider two modifications:

1. in the first one, the controlled phase is estimated not only when changing sensing time, but before each Ramsey measurement (*"full-adaptive" protocol*). The improvement achieved with this modification can be observed in Supplementary Figure S3, where we compare simulations for controlled phase updated only when changing sensing time and before each Ramsey. In the left plot, we compare the sensitivity, for increasing number of measurements $N$ ($N = 2..10$) in the case ($G = 3, F = 0$). Both protocols scale better than the standard quantum limit only for the first few data-points (until $N > 4$). However, the absolute sensitivity of the full-adaptive protocol is a factor two better. In the central plot, the same curves are displayed for ($G = 3, F = 5$). For these parameters, Heisenberg-like scaling is maintained until the coherence time limit is reached. Again, the full-adaptive protocol is better than the limited-adaptive for all $N$. In the right plot, we show the minimum achieved sensitivity for both protocols, as a function of $F$. In all cases the full-adaptive protocol outperforms the protocol which updates the optimal phase only when changing the sensing time.

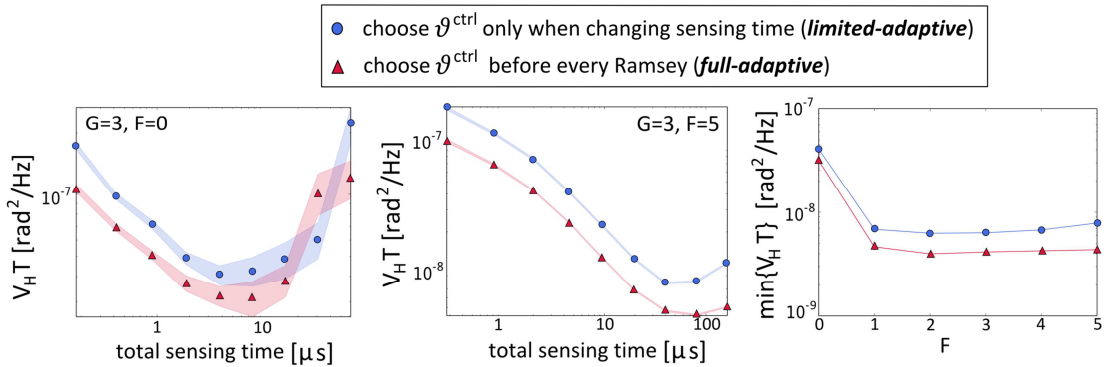

**Supplementary Figure S3.** *Adaptive protocols: simulation results comparing sensitivities obtained when updating the controlled phase only when changing sensing time (blue) and updating it before each Ramsey (red). We assume perfect readout fidelity and $T_2^* = 5\mu s$. Shaded areas represent error bars corresponding to one standard deviation (bootstrap method).*

Simulations in Supplementary Figure S3 are performed assuming perfect readout fidelity, $G = 3$, $T_2^* = 5\mu s$. Additional simulations (not shown) for different values of $G$, or imperfect readout fidelity, confirm the improvements gained by the full-adaptive strategy.

---

**Supplementary Table 3:**
**Optimized adaptive protocol**

$\bar{\mu} = 0$ (init)
for n = 1 to N:
    $t_n = 2^{N-n}$
    $M_n = G + F(n-1)$

    for m = 1 to $M_n$:
        choose $\vartheta_{n,m}^{ctrl}$
        if ($\bar{\mu} = 0$) then
            $\vartheta_{n,m}^{incr} = u0\ [m, n]$
        else:
            $\vartheta_{n,m}^{incr} = u1\ [m, n]$
        $\bar{\mu} = $ Ramsey $(\vartheta = \vartheta_{n,m}^{ctrl} + \vartheta_{n,m}^{incr},\ \tau = t_n \tau_{min})$
        Bayesian_update $(\bar{\mu}, \vartheta = \vartheta_{n,m}^{ctrl} + \vartheta_{n,m}^{incr},\ \tau = t_n \tau_{min})$

---

2. the second modification was suggested by A. J. Hayes and D. W. Berry [PS-3]. They proposed a protocol where the detection phase of the Ramsey experiment is $\vartheta_{n,m} = \vartheta_{n,m}^{ctrl} + \vartheta_{n,m}^{incr}$. A phase increment $\vartheta_{n,m}^{incr}$, dependent only on the last measurement outcome, is added to the controlled phase $\vartheta_{n,m}^{ctrl}$. Such phase increment is obtained by numerically optimizing the final variance in frequency estimation for the specific experimental parameters, through a "swarm optimization" procedure [PS-3,PS-6, PS-7] and tabulated in the functions u0, u1.
In this strategy the phase increments can be modelled by a binary decision tree. The size and direction of the steps depends on the last detection result and the current values of $n, m$. The particle swarm optimization (PSO) algorithm is then used to determine the steps in such a way that the final phase variance is minimized. In the PSO algorithm [PS-7], the problem space (the set of phase increments in this case) is searched by a swarm of particles. Each particle, labelled by $i$ has a velocity ($v_{id}(t)$) and a position ($x_{id}(t)$) which are updated according to its current best position ($x_{id}$) and the best position of the entire swarm ($x_{gd}(t)$) as:

$$v_{id}(t+1) = \chi\{v_{id}(t) + c_g r_g [x_{gd}(t) - x_{id}(t)] + c_l r_l [x_{ld}(t) - x_{id}(t)]\}$$
$$x_{id}(t+1) = x_{id}(t) + v_{id}(t)$$
(Eq. S-E6)

Here $r_g$ and $r_l$ are uniform random numbers in the interval [0; 1], $\chi$, $c_g$ and $c_l$ are constants, $d$ is the dimension of the space and $t$ is the number of the iteration. In our simulations we used $\chi = 0.729$, $c_l = c_g = 2.05$ with 10 particles and 400 iterations.

The optimized adaptive protocol, described in Box-3, combines phase increments with update of the controlled phase before each Ramsey. A comparison between the minimum sensitivity achieved by the limited-adaptive, non-adaptive and optimized-adaptive protocols is reported in Supplementary Figure S4. We fix $G = 5$ and assume $T_2^* = 96\mu s$.
The optimized adaptive protocol appears to perform always at least as good as the best between the limited-adaptive and the non-adaptive protocols. For lower values of $F$, the non-adaptive protocol fails to reach HL-scaling, while both adaptive ones do. For higher values of $F$, both the non-adaptive and the optimized adaptive reach the minimum sensitivity. Note that, for a readout fidelity $F_0 = 0.88$, while the optimized adaptive protocol reaches HL-scaling for $(G = 5, F = 2)$ the non-adaptive one needs at least $F = 4$.

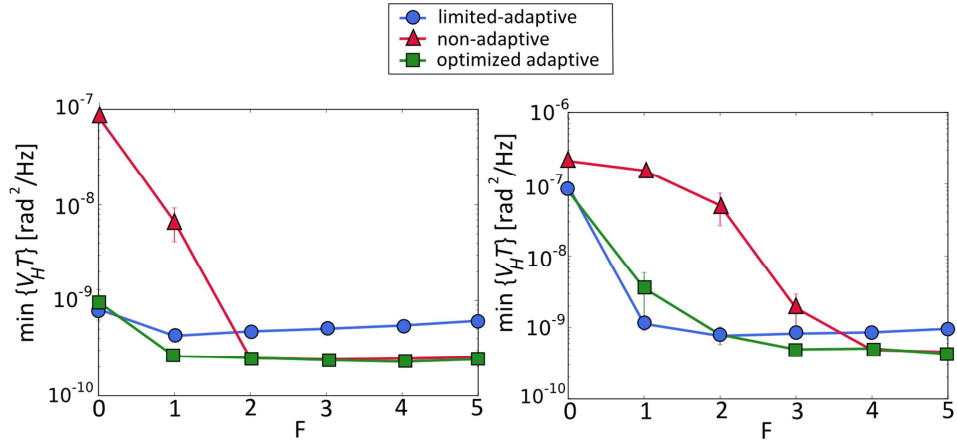

**Supplementary Figure S4.** *Simulations comparing the best achieved phase sensitivities for different protocols (left side, perfect readout fidelity $F_0 = 1$ – right side, $F_0 = 0.88$). We assume $G = 5$, $T_2^* = 96\mu s$. The plot on the right reports simulation results for the three protocols discussed in the main text, for experimental parameters and setting (as in main text).*

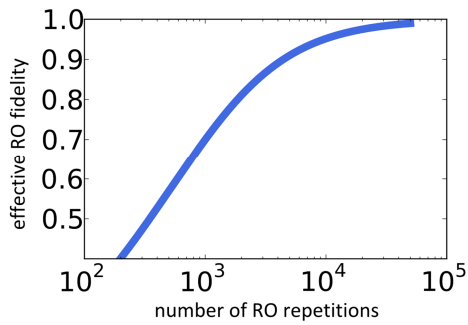

**Supplementary Figure S5.** *Effective readout fidelity as a function of readout repetitions, for measurements at room-temperature (no single-shot readout).*

**Application to room-temperature sensing.** At room-temperature, the spin-selective optical transitions within the zero-phonon line are not spectrally-resolvable and therefore single-shot spin initialization and readout by resonant optical excitation is not possible. Instead, the readout exploits the small difference in luminescence intensity by off-resonant optical excitation (in the green) when the electron spin is in $m_S = 0$, compared to $m_S = 1$.

In the following, we will use numbers from Nusran et al. [PS-5], that report $\alpha_0 = 0.031$ photons per shot when the electron is in $m_S = 0$, $\alpha_1 = 0.021$ photons per shot when the electron is in $m_S = 1$. Since one shot is not enough to perfectly distinguish between the two states, room-temperature experiments are repeated R times. The resulting signal level is then, respectively, $\alpha_0 R$ and $\alpha_1 R$. The contrast $C$ for a single repetition scales as [PS-8]:

$$\frac{1}{C} = \sqrt{1 + \frac{2(\alpha_0 + \alpha_1)}{(\alpha_0 - \alpha_1)^2 R}} \quad \text{(Eq. S-E7)}$$

This contrast $C$ is related to the fidelity with which the two states can be distinguished and, since luminescence detection is shot-noise limited, the error scales at the standard quantum limit as $R^{-1/2}$. Nusran et al. achieve a fidelity of 0.99, in their experiment [PS-5], by using 50000 readout repetitions per step. The achieved contrast as a function of readout repetitions is plotted in Supplementary Figure S5.

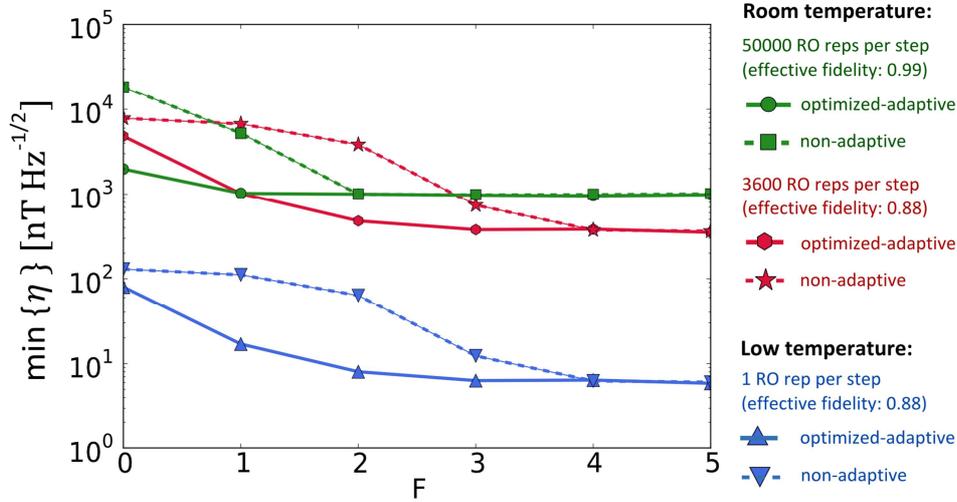

**Supplementary Figure S6.** *Simulations comparing the minimum magnetic field sensitivity (in nT Hz$^{-1/2}$) for the optimized adaptive and the non-adaptive protocols at room-temperature and low-temperature ($T_2^* = 96\mu s$, $N = 10$, $G = 5$).*

A contrast $C = 0.75$ can be achieved with $R = 1350$ repetitions, while $R = 3600$ repetitions are needed for $C = 0.88$, significantly less than the repetitions (50000) needed for almost perfect readout ($C = 0.99$).

In the simulations in Supplementary Figure S6, for consistency with previous results, we assume asymmetric readout fidelity ($F_1 = 0.993$, $F_0 = C + (1 - F_1)$), based on the contrast $C$ achieved with a given number of readout repetitions. The asymmetry of the readout fidelity can be controlled at will by choosing the threshold in photo-counts distinguishing $m_S = 0$ from $m_S = 1$.

Simulation results show that, at room temperature, the use of 50000 repetitions can achieve a sensitivity of $\eta \sim 1\mu T$ Hz$^{-1/2}$, either using the adaptive or non-adaptive protocols. However, using 3600 repetitions per step (with a lower effective readout fidelity), a better sensitivity $\eta \sim 0.4\mu T$ Hz$^{-1/2}$ can be reached. Moreover, for $F = 2$, the performance of the optimized-adaptive protocol with 3600 readout repetitions per step surpasses both the performance of the non-adaptive for the same conditions and the performance of the protocols with 50000 repetitions per step. For $F \geq 4$ adaptive and non-adaptive reach the same sensitivity: however, as discussed above and in the main text, a smaller value of $F$ allows a higher repetition rate of the estimation sequence.

This suggests the possibility that adaptive sensing, which reaches Heisenberg-limited scaling for a reduced number of measurements even in situation of lower fidelity, may be advantageous for room-temperature sensing, compared to non-adaptive protocols.

Simulations confirm the superior performance of the protocol in the case where single-shot readout is available, enabling sensitivities on the order of a few nT Hz$^{-1/2}$, as demonstrated experimentally by the data reported in the main text.

# II. Supplementary experimental details

(a) **Additional information about the experimental setup**.
*Sample.* We use an isotopically-purified diamond sample grown by Element Six Ltd ($^{13}$C concentration 0.01%) with a <100> –crystal orientation and study naturally occurring NV centres located 5–15μm below the surface. A solid immersion lens (SIL) is deterministically etched by focused-ion-beam (FEI Strata DB 235, 30 kV gallium ions) around a specific NV centre to increase collection efficiency. After milling the lenses, the sample is cleaned for 30 minutes in a boiling mixture of equal parts of perchloric, sulfuric and nitric acid. A single-layer aluminium-oxide antireflection coating, designed for best performance at a wavelength of 637 nm, is deposited on top of the sample [PS-8]. A 200 nm thick gold microwave strip line is defined along the lenses via electron beam lithography.

*Optics.* The sample is mounted on a stepper/scanner piezo stack (Attocube) in a Janis ST-500 flow cryostat with optical access and kept at a temperature of T = 8 K. At low temperatures, NV centres exhibit narrow zero phonon lines (ZPL) around 637 nm which we address resonantly with two lasers (New Focus Velocity tunable external cavity diode laser and Sirah Matisse DS dye laser with DCM dye). Each laser is locked to a wavemeter to ensure long-term frequency stability. In addition, yellow excitation at 575 nm from a frequency-doubled diode-laser (Toptica) is used for charge-state control (see Section IIc). All lasers are pulsed by acousto-optic modulators (Crystal Technologies).

We detect luminescence exclusively in the NV phonon sideband (650 - 750 nm), separated from the excitation light by dichroic filters (Semrock LPD01-633R). Photon detection is performed by avalanche photodiodes (Perkin Elmer SPCM).

*Spin control.* Microwave (MW) signals to drive the NV centre electron spin are generated by a Rohde Schwarz SMB100A source, with IQ modulation. To create the two π/2 pulses for each Ramsey experiment, the MW source output frequency is modulated (single-sideband modulation) by two pulses with rectangular envelope and 30 MHz carrier frequency. The first pulse is generated directly by an AWG (Arbitrary Waveform Generator, Tektronix AWG5014), while the second by a field-programmable logic array (FPGA). The FPGA receives the value for the phase $\vartheta$ chosen by the control microprocessor (ADwin Gold) and generates the modulation pulse with the proper IQ parameters, with timing triggered by the AWG. The value for the phase $\vartheta$ is specified as an 8-bit integer (256 levels), leading to a resolution of 1.4 degrees (0.025 radians).

Due to the hyperfine coupling to the host $^{14}$N spin, the electron spin transition is split into three frequency-separated lines. As discussed in Section IIb, these three lines are addressed separately by thee Ramsey experiments, realized by driving the electron spin at the three frequencies. This is achieved by an additional frequency modulation, imposed to the vector source by the AWG.

*Control microprocessor.* The microprocessor (Adwin Gold) manages the sequence of control pulses (optical and microwave) and counts the luminescence photon during spin readout. Moreover, it performs the Bayesian update of the probability density distribution $P(f_B)$ and chooses the proper controlled phase and phase increments.

The microprocessor code for Bayesian update minimizes the number of coefficients $p_k$ (Eq. S-E2) to be tracked and stored, to avoid exceeding the memory bounds of the microprocessor, and to optimize speed. We only use the coefficients which are known to be non-zero and contribute to the choice of the controlled phase $\vartheta$, neglecting the rest. Since the probability distribution is real ($p_k^* = -p_{-k}$), we can further reduce the computational requirements by only storing the coefficients for $k > 0$. Considering all this, the number of coefficients that needs to be processed, at each step $n$, is on the order of $M = G + F(n-1)$.

| N | initialization time [ms] | sensing time [ms] | readout time [ms] | computational time [ms] |
|---|---|---|---|---|
| 5 | 6.8 | 0.004 | 0.45 | 4.0 |
| 7 | 11.5 | 0.018 | 0.77 | 8.0 |
| 8 | 14.4 | 0.035 | 0.96 | 10.8 |
| 9 | 17.6 | 0.071 | 1.17 | 13.9 |
| 10 | 21.0 | 0.140 | 1.40 | 17.6 |
| 12 | 28.8 | 0.573 | 1.92 | 26.8 |

**Supplementary Table 4. Temporal budget of the estimation protocol.** *Total time, measured by the internal microprocessor clock, spent by the optimized-adaptive protocol in different tasks within the whole estimation sequence. The computational time (i.e. the time spent by the processor in performing the Bayesian update), is similar to that spent on spin initialization. Given that initialization and Bayesian update can be performed simultaneously, the computational time represents no additional overhead.*

In the case (G=5, F=2), the time spent by the microprocessor in the Bayesian update after each Ramsey experiment increases linearly from 80μs (for $n=2$) to 190μs (for $n=12$). This time is comparable with the spin initialization duration (150μs). In table 1 we show the total times associated with sensing, initialization, and computation of the Bayesian estimate. While in this work we performed the initialization and the Bayesian estimate sequentially, both operations can be performed simultaneously. In this way the real-time Bayesian estimation, a crucial prerequisite for the adaptive technique, does not add any temporal overhead to the protocol. In future implementations, the Bayesian estimation could be implemented with a dedicated FPGA, instead of a general-purpose microprocessor, which would allow a further reduction of the calculation time.

(b) **Microwave pulses and coupling to the $^{14}$N spin.** When using the electron spin of the NV center as a sensor in a Ramsey interferometry experiment, the coupling to its host $^{14}$N nuclear spin has to be taken into account. The hyperfine interaction ($\hat{H}_{hf} = A_{hf}\hat{S}_z\hat{I}_z$, neglecting small off-diagonal terms) effectively splits the electron spin $m_S = 0$ to $m_S = -1$ transition in three lines (Supplementary Figure S7b). Although these three lines can be addressed simultaneously by selecting a Rabi frequency larger than the coupling strength, the phase acquired during free evolution will depend on the state of the $^{14}$N spin. This creates ambiguity in the frequency estimation protocol, since the aim is to sense only the change in energy levels introduced by the Zeeman shift induced by the applied magnetic field, not the coupling to the $^{14}$N spin.

To circumvent this problem, we perform three sequential Ramsey sequences where, in each sequence, the microwave pulses are resonant with one of the three $m_S = 0$ to $m_S = -1$ transitions and the acquired phase only depends on the Zeeman shift. We choose the Rabi frequency (140 kHz) such that the pulses in each sequence only address selectively one of the three transitions. In Fig S7a, we show that we can perform Rabi oscillations selectively on the Nitrogen spin. Here the microwave pulses only drive the electron spin $m_S = 0$ to $m_S = -1$ transition if they are on resonance, thus for the $^{14}$N spin in a mixed state, a contrast of $1/3$ is expected. Full contrast is recovered when the three pulses are applied sequentially (Supplementary Figure S7c).

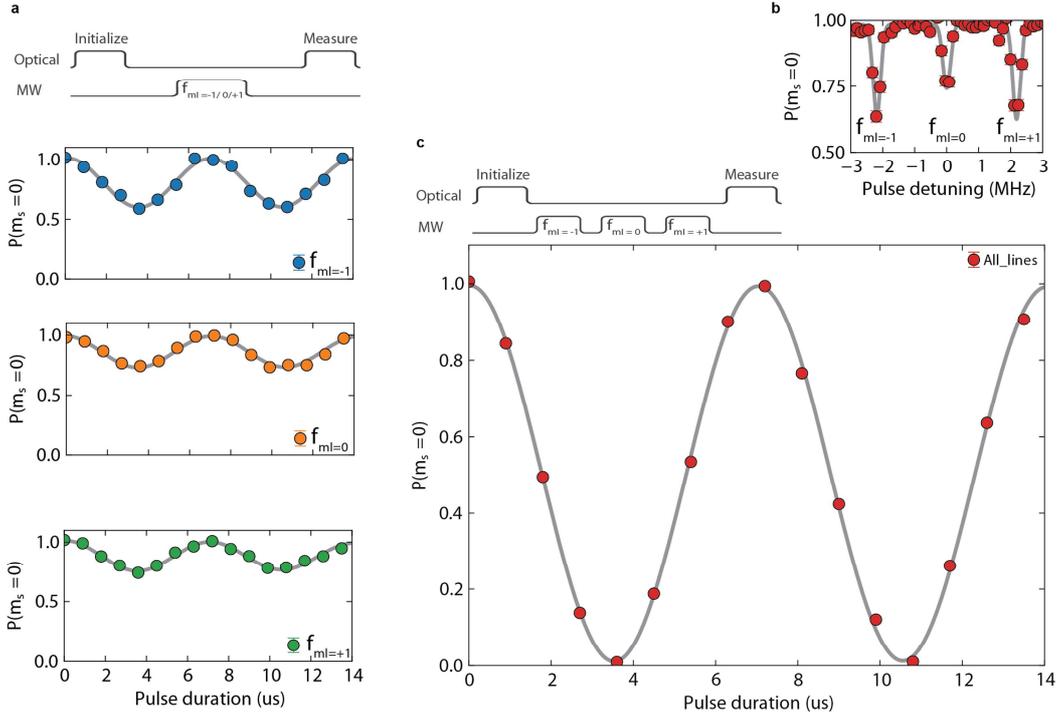

**Supplementary Figure S7. *Electron spin driving. a)*** *Rabi oscillations of the electron spin conditional on the state of the nitrogen spin ($^{14}$N , I=1). We tune the frequency of the microwave pulses in resonance with one of the three $m_S = 0$ to $m_S = -1$ transitions, corresponding to the nitrogen spin being either in $m_I = -1, 0$ or $+1$ (top, middle bottom) and vary the length of the pulse. From a sinusoidal fit (grey line) we find Rabi frequencies of (144, 140 and 142 ± 2 kHz) respectively **b)** Energy level spectrum for the electron $m_S = 0$ to $m_S = -1$ transition. We initialize the electron spin in $m_s$ =0 and then vary the frequency of a microwave pulse with fixed length. The pulse detuning is with respect to a reference frequency of 2.845334 GHz. The spectrum shows three lines owing to the hyperfine interaction with the $^{14}$N spin with $|A_{hf}|$ = 2π x (2.185 ± 0.006) MHz. **c)** Rabi oscillation of the electron spin unconditional on the state of the nitrogen spin. We apply three sequential microwave pulses each on resonance with one of the hyperfine lines. From the sinusoidal fit (grey line) we find a Rabi frequency of (142 ± 3) kHz.*

We note that this method requires the electron transition energies and therefore the static magnetic field to be known within the bandwidth of the pulses (~ 140 kHz). This is not a problem for our implementation, where the effect of an external field is implemented as an artificial detuning by adjusting the phase of the final π/2 pulse. When estimating a real magnetic field possible solutions would be to initialize the nitrogen spin, adjust the frequency estimation protocol to allow for sensing of multiple frequencies with fixed offset or adjust the interaction times such that the phase acquired during free evolution is independent of the state of the nitrogen spin ($2\pi = \tau A_{hf}$)

(c) **NV charge state and optical resonance pre- and post-selection.** Due to environmental charge noise, the optical transitions of the NV centre shift in frequency on a range larger than the linewidth. Moreover, resonant excitation can result in ionization of the NV$^-$ charged state into the neutral NV$^0$ state.

Before each estimation sequence, we check that the centre is in the NV$^-$ state, with optical transitions resonant with the excitation lasers. We turn both the initialization and readout lasers (on transitions E' and Ey, respectively) for 150 μs and count luminescence photons. Only if the luminescence photo-counts are larger than a given threshold (40 counts), the estimation sequence is started (*charge and optical resonance pre-selection*). We take the absence of luminescence photo-counts as an indication that the centre is ionized into the $NV^0$ state: the correct charge state is restored by resonant optical excitation of the $NV^0$ transition at 575 nm.

An estimation sequence can consists of a large series of Ramsey experiments, with spin initialization and readout. Ionization of the defect or large frequency shifts of the optical transitions during the sequence results in incorrect spin readout and errors in the magnetic field estimation. Therefore, we perform a new check of the charge and optical resonance conditions at the end of the estimation sequence and consider it as a valid estimation only if more than 10 luminescence photo-counts are detected (*charge and optical resonance post-selection*).

In the histograms in Supplementary Figure S8, we report an example of the number of rejected runs for 252 repetitions of the estimation sequence. While the average number of rejections in the post-selection process is around 50%, we have a consistent fraction of events (75/225) with no rejections, and other runs with 80% failure rate. This large spread is due to the fact that the data was taken in long automated measurement session during nights, with infrequent optimizations of the experimental parameters (like spatial overlap of laser beams on the NV centre). We believe that the percentage of rejected runs can be drastically reduced by optimizing the experimental settings and procedures.

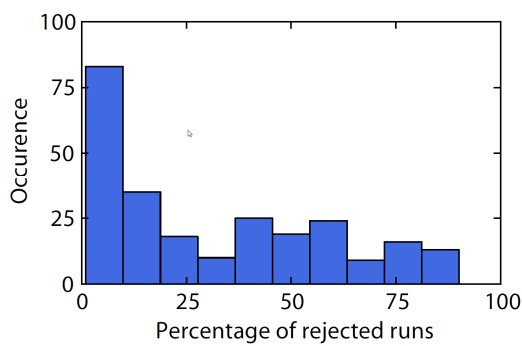

**Supplementary Figure S8.** *Histograms of the percentage of rejected runs in charge and optical resonance post-selection.*

**Supplementary References**